\documentclass[11pt,letterpaper]{article}

\usepackage{iccv}
\usepackage{times}
\usepackage{epsfig}
\usepackage{graphicx}
\usepackage{tabularx}
\usepackage{setspace}
\usepackage{amsmath}
\usepackage{amssymb}
\usepackage{caption,subcaption}
\usepackage{mwe}
\DeclareMathOperator*{\argmax}{arg\,max}

\usepackage[pagebackref=false,breaklinks=true,letterpaper=true,colorlinks,bookmarks=false]{hyperref}

\iccvfinalcopy 

\begin{document}

\centerline{\Large \bf Edge Projection-Based Adaptive View Selection for Cone-Beam CT}

\medskip
\centerline{Jingsong Lin, Singanallur Venkatakrishnan, Gregery Buzzard, Amir Koushyar Ziabari, Charles Bouman
\footnote{
Lin and Bouman are with Electrical and Computer Engineering, Buzzard is with Mathematics, all at
 Purdue University, West Lafayette,
 IN, USA. (E-mail: lin1311, buzzard, bouman@purdue.edu).
 Venkatakrishnan, and Ziabari are with the Multi-modal Sensor Analytics Group, Oak Ridge National Lab, Oak Ridge, TN, USA 37831 (Email: venkatakrisv, ziabarik@ornl.gov)

Copyright Statement: Research sponsored by the US Department of Energy, Office of Energy Efficiency and Renewable Energy, Advanced Materials \& Manufacturing Technologies Office (AMMTO), under contract DE-AC05-00OR22725 with UT-Battelle, LLC. 
The US government retains and the publisher, by accepting the article for publication, acknowledges that the US government retains a nonexclusive, paid-up, irrevocable, worldwide license to publish or reproduce the published form of this manuscript, or allow others to do so, for US government purposes. DOE will provide public access to these results of federally sponsored research in accordance with the DOE Public Access Plan (http://energy.gov/downloads/doe-public-access-plan).

}
}

\medskip
\medskip

\begin{abstract}
Industrial cone-beam X-ray computed tomography (CT) scans of additively manufactured components produce a 3D reconstruction from projection measurements acquired at multiple predetermined rotation angles of the component about a single axis. 
Typically, a large number of  projections are required to achieve a high-quality reconstruction, a process that can span several hours or days depending on the part size, material composition, and desired resolution. 
This paper introduces a novel real-time system designed to optimize the scanning process by intelligently selecting the best next angle based on the object’s geometry and computer-aided design (CAD) model.
This selection process strategically balances the need for measurements aligned with the part’s long edges against the need for maintaining a diverse set of overall measurements. 
Through simulations, we demonstrate that our algorithm significantly reduces the number of projections needed to achieve high-quality reconstructions compared to traditional methods. 

\end{abstract}

\noindent

\section{Extended Summary}

\subsection{Introduction}
Industrial X-ray cone-beam (CB) CT is widely used for non-destructive characterization of parts in additive manufacturing. 
A typical scan involves rotating a part about a single axis, measuring projections at each orientation of the object and processing all the measurements at the end of the scan using an analytic reconstruction algorithm like Feldkamp, Davis, and Kress (FDK) \cite{FDK} to obtain a 3D reconstruction. 
Typically, a large number of projections with sufficiently high SNR are acquired so that algorithms like FDK can produce high-quality reconstructions.
For industrial CBCT systems, it is typical to use detectors of size approximately $2000 \times 2000$ pixels, which roughly corresponds to requiring about $2000$ projection measurements for a high quality FDK reconstruction. 
Depending on the density of the sample, its size and the desired resolution, the total scan time can range from a few hours to even a day thereby restricting the use of CBCT to the characterization of a small number of parts. 

In order to reduce the scan time of CBCT systems while preserving image quality, researchers have used conventional sparse scanning protocols (using a uniform subset of the typical dense set of orientations) combined with advanced model-based iterative reconstruction (MBIR) and deep-learning (DL) algorithms \cite{simurgh} demonstrating that it is possible to reduce scan times dramatically while preserving the image quality. 
However, these workflows rely on measurements from a  fixed uniform sub-set of the typical orientations, and do not adapt to the specifics of the unique geometry of the part or the end goal of the CT scan (say to extract porosity or measure object metrology). 

In this paper, we envision a real-time X-ray CBCT system where projections are acquired and used for intermediate 3D reconstructions, and a decision is made on how to orient the part for the next measurement with the goal to reduce the total number of projections for a high-quality reconstruction. 
Our work is inspired from the work in \cite{edge alignment} which developed a similar idea for parallel beam hyper-spectral neutron CT. 
The core idea of the work in \cite{edge alignment} was an algorithm to select a new orientation based on maximizing an objective function that encourages making new measurements along the longest edges in the part while ensuring measurement diversity. 
We use the same heuristic to select new angles as in \cite{edge alignment}, but introduce fundamental changes to be able to use it for cone-beam CT.

\subsection{Proposed Algorithm}
In \cite{edge alignment}, a new orientation for a parallel beam CT system is selected by maximizing a function that balances two sets of terms -  an edge alignment score and an angle dispersion score. 
Specifically, the edge alignment score is computed as a function of the number of overlapping pixels of the edges in the reconstructed image and a binary mask generated based on the view angle direction. 
The angle dispersion score is computed using the difference between view angles with appropriate wrapping performed so that angles that are $180$ degrees apart are $0$ distance apart. 
There are a few challenges with directly using this method for our X-ray CBCT system.  
First, the edge alignment score only shows how well a fixed view direction aligns with the edges, which cannot accurately capture the fact that for a CBCT system each projection measures rays oriented in a range of different directions. 
Second, in CBCT, the view angles $\theta$ and $\theta+180^o$ are not symmetric but depend on the geometry of the CT scan; hence the metrics in \cite{edge alignment} cannot be directly used. 
Third, the computation time is large, which can be a challenge when working with large 3D volumes typically encountered in industrial CBCT. 
To solve these issues, we propose the following two new functions.
 
The first is the edge alignment function $I(\theta,x;\beta)$ which evaluates the alignment  of a view corresponding to an angle $\theta$ with the edges in the current reconstruction $x$. 
To compute $I(\theta,x;\beta)$, we begin by computing the binary edge image $\tilde x$ using the Canny edge detection algorithm \cite{canny} applied to the input image $x$. 
Next, we compute the tomographic forward projection of the binary edge image
\begin{equation}
    \tilde y_\theta=A_\theta \tilde x
\end{equation}
where $A_\theta$ represents the forward projection matrix for view $\theta$. 
If a view $\theta$ is better aligned with the edges in the image, it will have longer path length through the edge image. 
Therefore, the larger values in $\tilde y_\theta$ indicates the view $\theta$ is better aligned with the edges. 
Finally, the output of edge alignment function $I(\theta,x;\beta)$ is computed by the weighted sum of $\tilde y_\theta$
\begin{equation}
  I(\theta,x;\beta) = \sum_{s\in S} w_s \tilde y_{\theta,s}
\end{equation}
where $S$ is the set of all pixel indices in $\tilde y_\theta$, and $w_s$ is calculated using the softmax function
\begin{equation}
  w_s = \dfrac{e^{\beta \tilde y_{\theta,s}}}{\sum_{k\in S}e^{\beta\tilde y_{\theta,k}}} 
\end{equation}

The second function, the angle dispersion function $D(\theta,\Theta_{n}^\ast;x_{CAD},\gamma)$, evaluates how close a given view angle $\theta$ is to elements in the set of previously selected views with angles $\Theta_{n}^\ast=\{\theta_1^\ast,\cdots,\theta_{n}^\ast\}$. 
This function encourages diversity in the selected views. 
The angle dispersion score is computed as
\begin{equation}
    D(\theta,\Theta_{n}^\ast;x_{CAD},\gamma)=\exp\left\{-\gamma \sum_{j=1}^{n}\dfrac{1}{d(\theta,\theta_j^\ast;x_{CAD})}\right\}    
\end{equation}
where $\gamma$ controls the decay rate. 
The pairwise distance $d(\theta_i,\theta_j;x_{CAD})$ is computed as 
\begin{equation}
    d(\theta_i,\theta_j;x_{CAD})= \dfrac{1}{|S|}\left\Vert BP(y_{\theta_i},\theta_i) - BP(y_{\theta_j},\theta_j) \right\Vert_1
\end{equation}
where $y_\theta$ is computed by the linear forward projection of the CAD model of the part: $y_\theta = A_\theta x_{CAD}$,  
$BP(y_{\theta},\theta)$ is the unfiltered back projection of $y_\theta$,
and $|S|$ is the number of voxels in $x_{CAD}$.

$d(\theta_i,\theta_j;x_{CAD})$ uses both projection and CT scan geometry information. 
Compared to the angular distance used in \cite{edge alignment}, this distance metric is more robust as it can be used for a variety of cone-beam CT scan geometry. 
Specifically, $d(\theta,\theta+180^o;x_{CAD})$ will adaptively change based on the CT scan geometry, making it a more accurate distance measure between different views.

In conclusion, our proposed algorithm leverages the edge alignment function $I(\theta,x;\beta)$ and the angle dispersion function $D(\theta,\Theta_{n}^\ast;x_{CAD},\gamma)$ to sequentially select optimal views. 
In each iteration, the $n^{\text{th}}$ newly selected view $\theta_{n}^\ast$ is determined by numerically solving the following optimization problem
\begin{equation}
  \theta_{n}^\ast = \argmax_{\theta \in \Theta \setminus \Theta_{n-1}^\ast} \{ \lambda_n I(\theta,x_{CAD};\beta) + (1-\lambda_n)I(\theta,\hat{x}_{n-1};\beta) + D(\theta,\Theta_{n-1}^\ast;x_{CAD},\gamma)\}
  \label{eq:6}
\end{equation}
where $\hat{x}_{n-1}$ is the reconstruction obtained from all the measurements from $1$ to $n-1$. 
As seen in the above equation, we utilize both the static edge alignment score using the CAD model and the dynamic edge alignment score using the current reconstruction volume. 
They combine to provide more accurate edge alignment information in all stages for selecting views. 
The parameter $\lambda_n$ is selected as a linear decay function of $n$. 
The purpose is to rely more on CAD model information in the early stage and rely more on the reconstruction image information in the later stage.

\subsection{Experiment Results}
To evaluate the performance of our proposed algorithm, we examine it by applying it to select the optimal set of views from a simulated data set. 
The ground truth and the CAD drawing are given. 
Both consist of $200$ slices, each with $200 \times 200$ pixels. 
The ground truth is similar to the CAD drawing but includes extra artificial pores inside the object. 
For a given view angle, the synthetic projection is generated from the ground truth using the polychromatic X-ray source model with additive noise. 
The projection is then linearized to reduce the beam-hardening effect.
Following this, MBIR is used, which is implemented in pyMBIR \cite{pymbir}, for image reconstruction.

\begin{figure}
     \centering
     \begin{subfigure}[b]{0.4\textwidth}
         \centering
         \includegraphics[width=\textwidth]{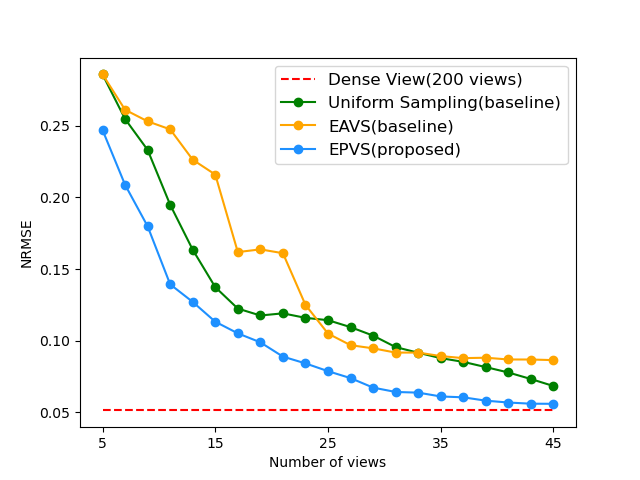}
         \caption{NRMSE plot}
         \label{fig:nrmse}
     \end{subfigure}
     \begin{subfigure}[b]{0.4\textwidth}
         \centering
         \includegraphics[width=\textwidth]{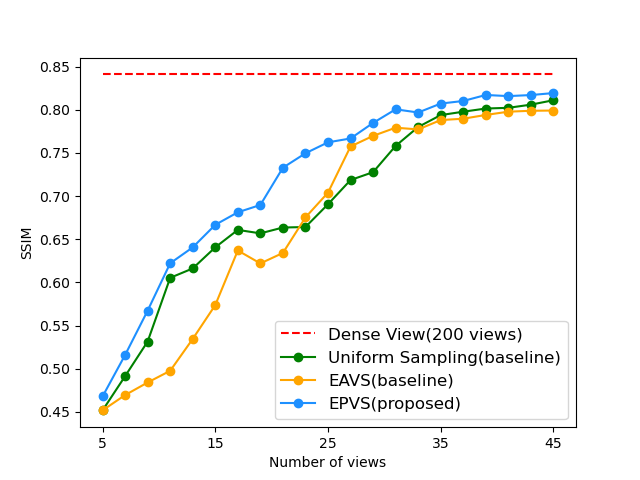}
         \caption{SSIM plot}
         \label{fig:ssim}
     \end{subfigure}
    \caption{Comparison of EPVS with other baseline methods (uniform sampling and EAVS) on reconstruction results during the process of selecting views. 
    NRMSE results are shown in (a), and SSIM results are shown in (b). 
    In these figures, the dashed lines show the NRMSE and SSIM results using 200 uniform sampling views. 
    All three algorithms start with $5$ views. 
    Both baseline methods are initialized with uniform sampling views, while EPVS is initialized by solving equation \ref{eq:6} with $\lambda_n=1$. 
    EPVS demonstrates better initialization by utilizing the information from the CAD drawing. 
    Additionally, EPVS converges faster compared to the baseline methods.}
    \label{fig:metrics}
\end{figure}

\begin{figure}
\centering
\includegraphics[height=6.5cm]{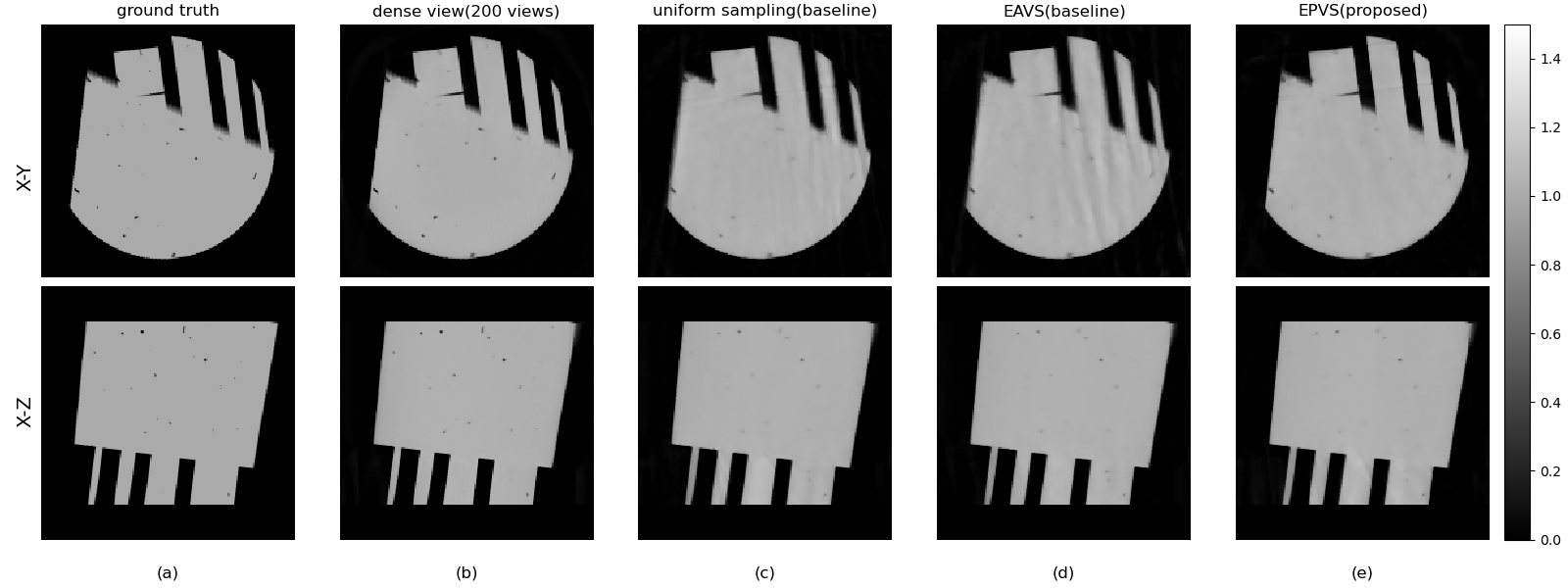}
\caption{Illustration of cross-sections from the  ground truth and  reconstructions ($200 \times 200 \times 200$ voxels) obtained using different view-selection algorithms. 
The first row shows XY cross-section and the second row shows XZ cross-section. 
(a) Ground truth. 
(b) Dense-view reconstruction from $200$ simulated measurements. 
(c) Reconstruction using 35 views and uniform sampling (baseline method). 
(d) Reconstruction using 35 views and EAVS (baseline method in \cite{edge alignment}). 
(e) Reconstruction using 35 views and EPVS (proposed method).
Notice that our method is free of streaking artifacts and is qualitatively closest to the ground truth compared to other algorithms.}
\label{fig:recon}
\end{figure}

Using the same experimental setup, we compare our proposed algorithm, edge projection-based view selection (EPVS), with two baseline algorithms: uniform sampling and edge alignment-based view selection (EAVS) as described in \cite{edge alignment}. 
In Figure~\ref{fig:metrics}, we compare the normalized root mean square error (NRMSE) and structural similarity index measure (SSIM) using the three different algorithms.
Figure~\ref{fig:recon} shows the reconstruction images using 35 selected views.

As shown in Figure~\ref{fig:nrmse}, EPVS has a faster convergence speed (i.e. it needs fewer views to achieve the same reconstruction performances achieved by the baseline algorithms). 
Also, the NRMSE curve for EPVS is uniformly below the ones using baseline algorithms which indicates our algorithm provides improvement for any number of views in our experimental setup. 
Finally, Table~\ref{table:computation time} compares the average time for selecting one view (excluding the reconstruction time) using EAVS and EPVS. 
The comparison indicates EPVS requires significantly less computation time compared to EAVS.

\begin{table}[!t]
\centering
\begin{tabular}{|m{.3\textwidth}| m{.3\textwidth}|}

\hline
 \center{EAVS(baseline)} & \center{EPVS(proposed)} \tabularnewline
\hline
\center{$259.03s$}  & \center{$11.20s$}\tabularnewline
\hline
\end{tabular}
\caption{Comparison of the average time for selecting one view (excluding the reconstruction time) using EAVS(baseline) and EPVS(proposed).
Notice that EPVS requires significantly less computation time compared to EAVS.}
\label{table:computation time}
\end{table}

\end{document}